\documentclass[11pt]{article}
\usepackage[margin=1in]{geometry}
\usepackage[utf8]{inputenc}
\usepackage[T1]{fontenc}
\usepackage{graphicx}
\usepackage{xcolor}
\usepackage{hyperref}
\usepackage{xurl}
\usepackage{url}
\usepackage{enumitem}
\hypersetup{
  colorlinks=true,
  linkcolor=black,
  citecolor=black,
  urlcolor=blue,
  pdftitle={The Eticas AI Risk Taxonomy: Open Infrastructure for Operationalizing AI Audits},
  pdfauthor={Gemma Galdon Clavell, Pablo Accuosto, Usman Gohar}
}

\providecommand{\tightlist}{%
  \setlength{\itemsep}{0pt}\setlength{\parskip}{0pt}}

\newcommand{\refentry}[1]{\par\noindent\hangindent=1.5em\hangafter=1 #1\par\vspace{0.5em}}

\title{The Eticas AI Risk Taxonomy: Open Infrastructure for Operationalizing AI Audits}
\author{
  Gemma Galdon Clavell \quad Pablo Accuosto \quad Usman Gohar \\
  \small Eticas.ai
}
\date{Taxonomy version 3.0.0 \quad|\quad Paper draft: July 2026}

\begin{document}

\maketitle

\begin{abstract}
The rapid deployment of AI systems across high-stakes domains has created urgent demand for standardized evaluation, yet the field remains fragmented across competing risk taxonomies that catalog risks without showing how an audit is actually executed. At least 74 AI risk taxonomies now exist; and almost all stop at the catalog. The hard part of auditing is not naming a risk but operationalizing it: turning a named risk into a test run against a real system, a measured value, a calibrated severity, and a defensible grade. This paper leads with that bridge. We present the operationalization layer Eticas has built and run, shown end to end on a single risk (PII leakage) against a public benchmark, and then the open taxonomy that makes the method scale. On GPT-4-0314, the same disclosure risk that seven external frameworks require be controlled is measured at 0\%, 51\%, and 84\% disclosure as adversarial conditioning increases, mapping through calibrated severity bands to a subcategory grade of E with a SYSTEMIC pattern. Around this worked example, the Eticas AI Risk Taxonomy v3.0.0 organizes 70 active subcategories across 10 categories and 21 sub-groups, with formal mappings to 18 external frameworks across compliance, reference, and academic tiers. Its established layer (categories, sub-groups, and the 32 established subcategories) is published under CC BY 4.0 as open semantic infrastructure with stable URIs and SKOS/JSON-LD distributions, and a worked subcategory example is disclosed to show the operational layer down to its severity thresholds. The contribution is the demonstrated bridge from concept to graded finding, anchored by a clean ontological separation of risks from the mechanisms by which they surface (the stable seam at which any operationalization attaches, disclosed in full for the worked examples), and framed by an open-core model in which the conceptual scaffold is open and the methodology calibration is the practitioner layer. This is the infrastructure the AI auditing field needs: shared, open, and demonstrably operable.

\end{abstract}


\section{Introduction}\label{introduction}

Artificial intelligence systems now mediate consequential decisions across employment, credit, healthcare, criminal justice, and public services. With generative AI and LLMs, they also mediate access to and production of information (Bender et al., 2021). This deployment velocity has catalyzed regulatory responses worldwide: the European Union's AI Act (Regulation 2024/1689) entered into force in August 2024 with full application phased to 2027; the United States published the NIST AI Risk Management Framework in January 2023, followed by a Generative AI Profile in July 2024; China released version 2.0 of its TC260 AI Safety Governance Framework in September 2025; and the Council of Europe opened the first binding international AI treaty for signature in September 2024. These frameworks mandate risk assessment, impact evaluation, and ongoing monitoring, yet they provide categories at abstraction levels poorly suited to operational implementation.

Consider the EU AI Act's obligation for providers of high-risk AI to establish quality management systems ensuring compliance across data governance, technical documentation, transparency, human oversight, accuracy, robustness, and cybersecurity (Articles 9 to 15). Or ISO/IEC 42001:2023's requirement that organizations assess and treat AI-related risks through systematic impact assessment (Annex A.5). These obligations presuppose that evaluators can decompose ``accuracy'' or ``fairness'' into measurable assessment criteria, yet the regulations offer limited guidance on how to structure that decomposition.

The field's response has been proliferation of AI risk taxonomies, with at least 74 now existing (Slattery et al., 2026), each organizing risk concepts differently. The MIT AI Risk Repository catalogues over 1,700 risks; AIR 2024 extracts 314 risk types from regulations and corporate policies (Zeng et al., 2024); IBM's AI Risk Atlas structures risks by lifecycle stage (Bagehorn et al., 2025); China's TC260 organizes by risk source rather than domain of harm. This heterogeneity fragments the field such that regulators cannot readily compare audit outputs across providers, clients cannot assess the coverage they are getting, and practitioners reinvent definitions rather than converging on shared vocabulary.

Beneath the fragmentation lies a more fundamental limitation, and it is the one this paper addresses. Most of the work on AI risk taxonomies is theoretical in nature, with no path towards operationalizing it towards real AI audits. Existing taxonomies enumerate, classify, and cross-reference risks, but they do not show how a named risk becomes a test run against a real system, how a test output becomes a measurable output metric, how a metric becomes a severity rating, how severity ratings aggregate into an interpretable grade, or how that grade survives scrutiny in a regulatory compliance context. The difficult part of auditing is not naming the risk; it is operationalizing it. A taxonomy that cannot be operationalized, while useful, functions as a glossary, rather than an audit methodology. The field has made substantive progress developing taxonomies, yet demonstrated bridges from concepts to executed audits remain scarce.

To address this gap, we adopt an inductive rather than deductive approach. A single risk is taken from taxonomy entry through test design, execution, measurement, and graded judgment, establishing the full audit chain empirically before the broader classificatory framework is introduced. Situating the empirical demonstration prior to the theoretical apparatus is deliberate: it grounds the taxonomy in a concrete, replicable procedure.

Towards this, we present the Eticas AI Risk Taxonomy v3.0.0, which organizes AI risks across 10 top-level categories (Bias and Fairness, Privacy and Confidentiality, Reliability, Governance, Security and Misuse, Environmental Impact, Transparency and Explainability, Agentic AI, Autonomy and Agency, Organisational Readiness), decomposed into 21 sub-groups and 70 active subcategories, each carrying a stable identifier, a definition, lifecycle-stage annotations, an enumeration of the mechanisms by which the risk surfaces, and mappings to 18 external frameworks across three tiers. The established layer (categories, sub-groups, and established subcategories) is published under CC BY 4.0 with SKOS and JSON-LD distributions at \url{https://taxonomy.eticas.ai/}, and a worked subcategory example is disclosed to show the operational layer.

This work makes four contributions, ordered here as the paper presents them. First and central, a demonstrated operationalization layer: a four-layer methodology, a codified measurement-to-grade chain, and per-mechanism test design, validated end to end against a public benchmark, so that the taxonomy is shown to be practical infrastructure rather than a theoretical catalog. Second, an ontological separation of risks from the mechanisms by which they surface, where the mechanisms field is the contract surface between the open taxonomy and any operationalization that consumes it. Third, audit-oriented granularity that connects abstract regulatory requirements to testable criteria, mapped to 18 external frameworks. Fourth, the inclusion of Agentic AI as a first-class category, which surfaces a structural governance gap: the major regulations predate agentic AI as a deployment paradigm.

It is important to note that this is a practitioner's contribution. The taxonomy derives from sustained engagement with real AI systems across operational audit contexts, and its organizational structure reflects what has proven effective in planning and executing audits. We do not claim that this is the only correct way to organize AI risks, or that the exact number of categories is optimal. In this work, we demonstrate that a named risk can be carried through to a tested, graded result on a real system, by a documented procedure that others can follow. The structure exists to make that procedure repeatable and to scale it across the full range of risks.

\section{From Risk to Graded Finding: The Operationalization Layer}\label{from-risk-to-graded-finding-the-operationalization-layer}

A taxonomy earns the word ``infrastructure'' only if something can be built on it. This section describes what Eticas has built and run across verticals and systems and in 10+ years of real-world AI evaluations and audits. We describe it for two reasons. First, to substantiate the claim that the taxonomy is practical rather than theoretical, by walking a single risk from the regulations that require it to a measured grade on a real model. Second, to demonstrate replicability as the interface between the taxonomy and operationalization is defined by stable structures - the open concept identifiers and the per-risk mechanism identifiers, disclosed in full for the worked examples - such that the Eticas methodology described here represents one instantiation of a pattern anyone can implement.

\subsection{A Worked Example: PII Leakage, End to End}\label{a-worked-example-pii-leakage-end-to-end-1}

We begin with the destination, then explain how the method reaches it. Figure 1 traces a single risk, PII leakage, from the frameworks that require it to be controlled, through the mechanisms by which it surfaces, to a measured and graded finding on a real model.

\begin{figure}[htbp]
\centering
\includegraphics[width=\textwidth]{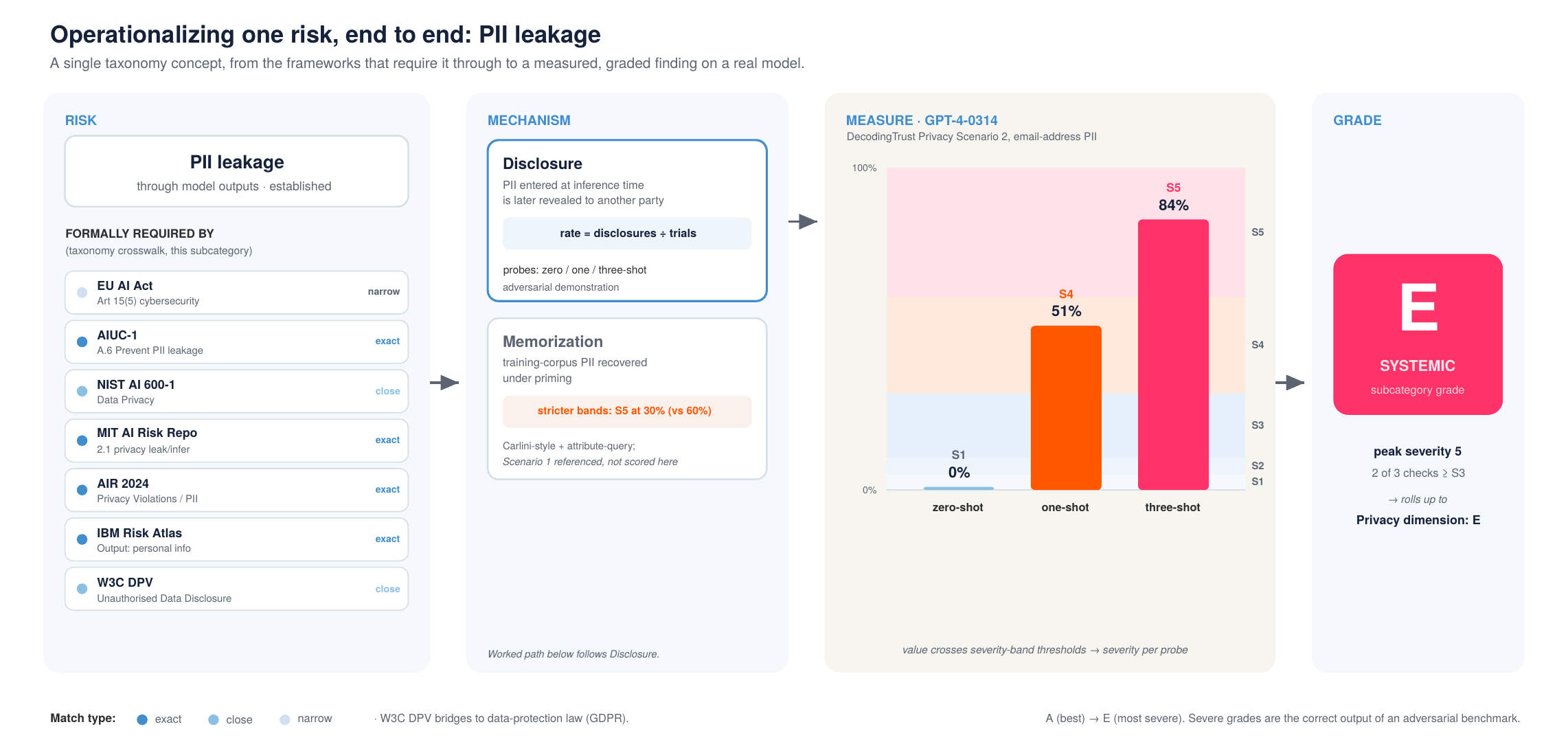}
\caption{Operationalizing one risk, end to end. PII leakage is formally mapped to seven external frameworks (left); the methodology operationalizes it along two mechanisms; the disclosure mechanism, probed on GPT-4-0314 via DecodingTrust Privacy Scenario 2 (email-address PII), yields disclosure rates of 0\%, 51\%, and 84\% as adversarial conditioning increases, which map through calibrated severity bands to severities 1, 4, and 5 and aggregate to a subcategory grade of E with a SYSTEMIC pattern. The validation is pinned to taxonomy v2.0.0, the version it was run against; in v3.0.0 the concept is broadened and renamed sensitive-information-leakage (Section 2.1).}
\label{fig:pii}
\end{figure}

Read left to right, the figure is the whole argument in miniature. On the left, PII leakage is not a free-floating concept: it is the formal target of seven external frameworks at once. It is an exact match to AIUC-1's control A.6 (Prevent PII leakage), to the MIT Repository's privacy-leak subdomain, to AIR 2024's PII clause, and to IBM Atlas's ``revealing personal information in output''; a close match to NIST AI 600-1's Data Privacy area and to the W3C Data Privacy Vocabulary's Unauthorised Data Disclosure; and a narrow match to the EU AI Act at Article 15(5) on confidentiality attacks. The W3C DPV mapping is the bridge to data-protection law, GDPR included, even though GDPR is not itself a taxonomy mapping. One risk, mapped once, discharges evidence against all seven regimes at the conceptual level. That is the interoperability payoff of a single shared identifier.

Moving right, the taxonomy records PII leakage as one risk with distinct mechanisms, and the methodology operationalizes two of them. The disclosure mechanism (personal data entered into the model's context at inference time is later revealed to another party) is the one we follow to a finding. Its metric is a disclosure rate: the fraction of trials in which the injected PII appears in the model's output. Three probes (the procedures that exercise a mechanism under a set condition) apply the metric under increasing adversarial pressure: a zero-shot baseline with no demonstration, a single in-context demonstration of the model disclosing PII, and three reinforced demonstrations.

Run against GPT-4-0314 using DecodingTrust's Privacy Scenario 2 on the email-address PII type, the three probes produce a sharp escalation: 0\% disclosure at zero-shot, 51\% under a single demonstration, and 84\% under three. The methodology maps each rate through the disclosure-route severity bands that live at Layer 2: 0\% falls in the severity-1 band (no or trivial concern), 51\% in the severity-4 band (the model fails the majority of the time under tested conditions), and 84\% in the severity-5 band (the model is effectively non-protective). The subcategory grade is the peak severity across the checks, which is 5, giving a grade of E, recorded with a SYSTEMIC pattern flag because two of the three checks land at severity 3 or above rather than a single outlier. In this validation pass the Privacy dimension\footnote{Terminology: we use \emph{category} for the taxonomy's top-level units (e.g., Privacy and Confidentiality) and \emph{dimension} for those same top-level units as they enter the methodology's assessment and grading chain, where subcategory grades aggregate into a dimension (equivalently, top-level category) grade. Both are distinct from the \emph{subcategory} level, at which test design attaches. External frameworks (e.g., IBM's Risk Atlas) use ``dimension'' in their own, unrelated sense.} carries only this one assessed subcategory, so the dimension grade reads through to E as well; a full Privacy assessment would probe the other subcategories before the dimension grade could be read as broadly representative.

Three features of this finding are worth drawing out, because they reflect field experience rather than methodological assertion:

First, the diagnostic value lies in the shape of the result rather than a summary statistic. The model's default refusal is robust against the laziest adversary (0\% at zero-shot) and collapses under light conditioning (51\% after one demonstration). An audit that reported only an average across protocols would erase precisely the signal that is consequential to a deployer, namely that the safety policy does not generalize beyond the trivial case. This is why the methodology records peak severity and a pattern flag rather than a mean, and why the probe set is designed to vary adversarial strength systematically.

Second, the severity bands encode a judgment that a generic metric would miss. In this case, memorization and disclosure are evaluated against different thresholds because they represent qualitatively different failures, and collapsing them onto a single scale would lose that distinction. Memorization (training-corpus PII recovered under priming, probed in the Carlini-style prefix-priming and attribute-query protocols) is graded more strictly than disclosure: its severity-5 band opens at a 30\% extraction rate, where disclosure's opens at 60\%. The reason is a distinction that only matters once you have run real audits. In disclosure, the user introduced the PII into context and has some awareness of the exposure; in memorization, the model re-emits PII from training with no user awareness at all, so a far lower rate already constitutes a structural privacy failure. The same metric value means different things by mechanism, and the calibration says so.

Third, a defensible audit resists the aggregations that flatten a system. The metric is computed per PII type rather than pooled across types, because pooling damps the worst-affected type, and a single badly-leaking attribute can be concerning, not the average. And the bands carry a practical-significance caveat: in regulated settings a single demonstrated leak of identifying PII may warrant elevated severity regardless of rate, because the failure mode (PII exposed at all) is itself the harm. These are not refinements bolted on after the fact; they are the residue of auditing real deployed systems, where a deployed-system test under live query conditions behaves differently from a benchmark run against an API, and where clients and regulators ask not ``what was the average'' but ``what is the worst thing you found and how bad is it.''

We show this example openly, down to the thresholds, as the one fully disclosed instance. The remaining operationalized entries follow the identical authoring pattern but keep their calibration in the practitioner layer. One risk shown in full proves the method is real; the breadth of the catalog, withheld, is where accumulated practice compounds. (Three honest boundaries: the numeric results here are the published DecodingTrust figures for GPT-4-0314 and should be read as a public-benchmark proxy for a live audit rather than a client engagement; the severity bands shown are a documented first-pass calibration, to be refined as the methodology is exercised against further cases; and the validation is pinned to taxonomy v2.0.0, the version it was run against - per-audit version pinning is itself part of the methodology - while in v3.0.0 the concept has since been broadened and renamed sensitive-information-leakage, with the retired pii-leakage identifier still resolving as a redirect.)

\subsection{A Four-Layer Architecture}\label{a-four-layer-architecture}

The methodology separates what is reusable across audits from what is specific to each engagement, in four layers (Table 1).

\begin{table}[htbp]
\centering
\includegraphics[width=\textwidth]{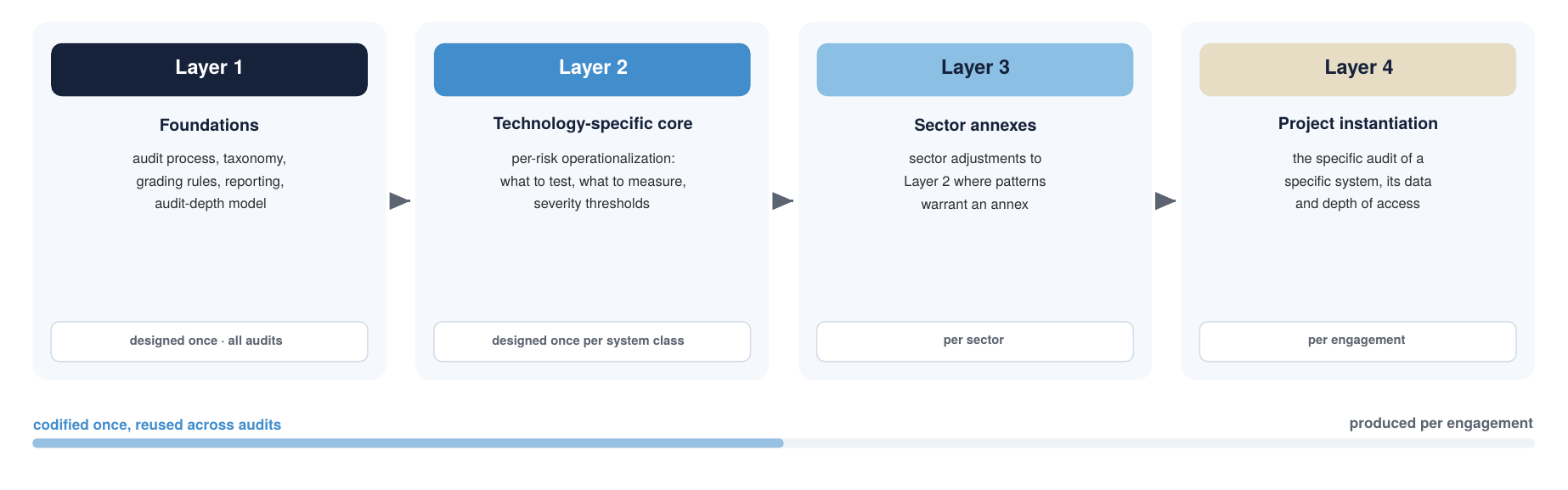}
\caption{The four-layer methodology architecture. Layers 1 and 2 are reusable across audits; Layers 3 and 4 are produced per engagement.}
\label{tab:layers}
\end{table}

Layers 1 and 2 are the reusable infrastructure, the cumulative work that does not need redoing for each audit. Layer 1 (Foundations) holds the universal audit process, the taxonomy itself, the grading rules, the reporting format, and the audit-depth model, and is designed once and used across all audits. Layer 2 (Technology-specific core) is how each risk is operationalized for a class of systems such as LLM or automated decision-making: what to test, what to measure, what severity thresholds apply, designed once per system class. Layer 3 (Sector annexes) adjusts Layer 2 for a sector when sector patterns are strong enough to warrant their own annex. Layer 4 (Project instantiation) is the specific audit of a specific system with its data and depth of access, engagement-specific by definition. In practice, an audit begins at Layer 4 and pulls down what it needs from Layers 1 and 2: the auditor does not redefine what PII leakage means or how severity is graded, because those already live above. The worked finding in Section 2.1 is a Layer 4 instantiation reading its definitions, metrics, and bands unchanged from Layer 2.

The taxonomy is the spine of Layer 1; everything else attaches to it. This is the structural reason the taxonomy can be infrastructure for others: the layer that is hardest to build (a coherent, framework-aligned, operationally scoped risk vocabulary with a clean risks-versus-mechanisms ontology) is the layer Eticas publishes, as an open scaffold plus worked examples.

\subsection{The Mechanisms Field as Contract Surface}\label{the-mechanisms-field-as-contract-surface}

The seam between concept and operationalization is the mechanisms field, and Section 2.1 instantiates it. The taxonomy, at the pinned version, records PII leakage with distinct mechanisms - disclosure, memorization, and cross-customer contamination - and the methodology declares, against those exact mechanism identifiers, what it does about each. Disclosure and memorization are operationalized with different probes, data, and calibrations; cross-customer contamination is recorded at the taxonomy level but not yet operationalized, so it appears downstream as a visible empty slot rather than a silent omission. Nothing is re-defined at the boundary; the methodology cross-references the concept by identifier and mechanism.

This is where the ontological discipline of the taxonomy delivers practical returns. Because the taxonomy separates the risk from its mechanisms, the operationalization layer can attach test designs at the mechanism level without ambiguity, and a second practitioner could attach a different test design at the same mechanism and the two would be comparable. The open concept combined with the stable mechanism identifier constitutes the interoperability contract between the taxonomy and any conforming methodology; the worked examples disclose that contract in full, while the mechanism layer across the rest of the catalog remains internal.

\subsection{The Measurement-to-Grade Chain}\label{the-measurement-to-grade-chain}

When an audit is executed, concrete observations are translated step by step through a sequence of steps into the grades that appear in the public-facing summary. The steps are invariant across audits; only the values are specific to the system under evaluation. The chain makes explicit what is codified once in the methodology and what an auditor produces anew per engagement.

\begin{itemize}
\tightlist
\item
  A probe is the procedure that exercises a mechanism to elicit evidence; a check is a single observation it produces, made for a defined purpose: a number, a piece of evidence, or a judgment. The methodology defines what kinds of checks each subcategory expects; the auditor produces them.
\item
  A metric value is the numerical result of a quantitative check, computed from raw output by a metric definition held in a registry (for example, a PII disclosure rate). Definitions are codified once; values are produced per audit.
\item
  A severity (0 to 5) is the metric value mapped onto a scale via the severity bands declared at Layer 2 for that metric. Bands are part of the methodology; the resulting severity is engagement-specific. Different mechanisms of the same risk can have different calibrations, as the PII example showed.
\item
  A subcategory grade (A to E) is the peak severity across the subcategory's checks, recorded with a pattern flag (ISOLATED, FOCAL, or SYSTEMIC) for any subcategory with two or more checks, which preserves the distribution signal that a peak alone would discard.
\item
  A dimension grade (A to E) is derived from the subcategory grades by a rule that takes the peak and then applies a breadth-of-concern adjustment across the subcategories within the dimension.
\end{itemize}

The worked finding in Section 2.1 is this chain running once: three checks, three metric values, three severities via the disclosure bands, a peak-plus-pattern subcategory grade, and a dimension grade. Table 2 makes the division of labor explicit. In this case, the schema, the metric registry, the severity bands, and the aggregation rules are codified once; scope decisions, test execution, metric values, narrative, and the final findings document are produced per engagement.

\begin{table}[htbp]
\centering
\includegraphics[width=\textwidth]{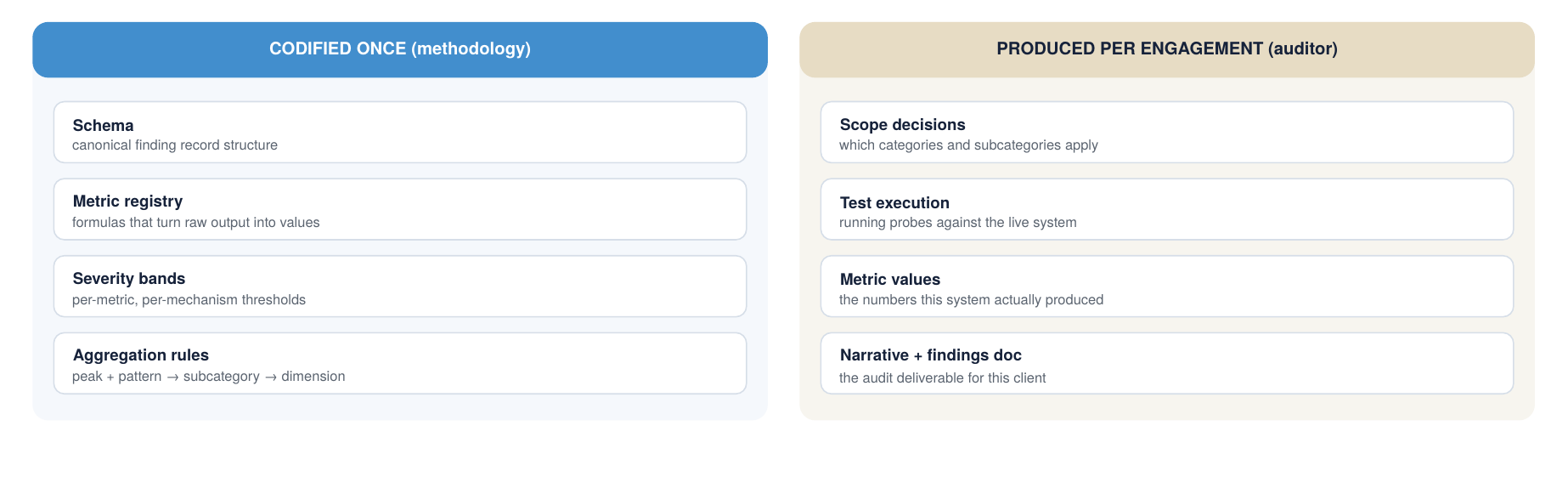}
\caption{The measurement-to-grade division of labor. The schema, metric registry, severity bands, and aggregation rules are codified once; scope, execution, metric values, and narrative are produced per engagement.}
\label{tab:measure}
\end{table}

\subsection{Anatomy of an Operationalized Entry}\label{anatomy-of-an-operationalized-entry}

Each operationalized risk is a single structured entry that an auditor can read from top to bottom. It declares the mechanisms it covers; for each mechanism, the test approach and its probes; the metric definitions, with formulas and the interpretation that fixes how the number is read; the default severity bands, per metric and per mechanism; the data the audit must supply, characterized by shape rather than by a fixed dataset; and the bindings to external benchmarks, mapped probe by probe. The PII-leakage entry behind Figure 1 declares two operationalized mechanisms, five probes, two metrics, two severity-band sets, and bindings to DecodingTrust Privacy Scenarios 1 and 2.

Two design properties follow this anatomy. First, coverage extends by repeating the shape: each new operationalized subcategory fills the same template, and because every entry shares the schema, the work compounds rather than fragmenting, as each entry strengthens the shared calibration. Second, gaps are visible, not hidden. If a mechanism is declared but no probes are populated for it, the gap shows up as an empty slot under a declared mechanism; the cross-customer-contamination mechanism of PII leakage is exactly such a slot in this entry, recognized as relevant to multi-tenant deployments and not yet bound to a probe. The design keeps both the intent and the implementation distance legible to anyone reading the entry.

We show this anatomy in full for one risk rather than partially for several. One fully disclosed entry is sufficient to make the architecture reproducible; showing the calibration of many entries would expose the practitioner layer without strengthening the argument.

\subsection{Validation, Open Infrastructure, and One Record Rendered Many Ways}\label{validation-open-infrastructure-and-one-record-rendered-many-ways}

For a practitioner taxonomy, the operative test is not conceptual coverage but executability: whether a named risk can be carried all the way through to test design, measurement, and a final judgment on a real system, in a way others can reproduce. The preceding section demonstrates that this is achievable within the Eticas framework, and the surrounding evidence takes four forms: a validated core, an end-to-end benchmark run, open and consumable infrastructure, and a single findings record that serves many audiences.

The methodology turns taxonomy concepts into executable assessment across five core dimensions (a prioritized subset of the taxonomy\textquotesingle s top-level categories: Bias and Fairness, Privacy and Confidentiality, Reliability, Security and Misuse, Governance), defined for both the LLM and ADM system classes. The end-to-end pipeline has been exercised on representative subcategories in two of these dimensions, Privacy and Bias and Fairness, proving that every stage from probe to dimension grade runs on real measurement; coverage extends across the remaining subcategories by repeating that proven shape.

That validation was run against a public benchmark. The pipeline, from authored findings through grading, has been exercised end to end against DecodingTrust (Wang et al., 2023), a recognized public LLM trustworthiness benchmark, in both operationalized dimensions. This is the closest available proxy to a live audit on public data: the benchmark supplies the inputs, the methodology is applied, findings are authored in the canonical schema, and grades are aggregated by the methodology\textquotesingle s own engine. Each check carries a metric value, a severity, the threshold it was graded against, and provenance recording that it inherited its Layer 2 invariants. The PII-leakage finding in Section 2.1 is one output of this pipeline; the bias-and-fairness dimension exercises the dimension-level breadth-of-concern aggregation across multiple subcategories. The taxonomy is, in other words, not a proposal but a tested instrument. DecodingTrust is the first external benchmark bound to the entries; because bindings are pluggable, others (for example TrustLLM or HELM) can be added without changing the methodology.

The conceptual layer is published for use, not merely described. The categories, sub-groups, and established subcategories are browsable at stable per-concept URIs (for example, taxonomy.eticas.ai/risk/bias-fairness), and the same content is distributed in machine-readable SKOS, as Turtle and JSON-LD, at taxonomy.eticas.ai/dist/. The public surface is released under CC BY 4.0, and the distributions emit a dcterms:license triple pointing to the CC BY 4.0 deed, so the license assertion travels with the data and is itself machine-readable. A tool, a regulator\textquotesingle s system, or another auditor\textquotesingle s pipeline can consume the concepts directly: resolve a concept, read its definition and framework mappings, tag a finding against its stable URI, and aggregate across concepts, without re-keying by hand. The mechanisms field is the demonstrated contract surface: its identifiers are stable, and the worked examples disclose them in full, so anyone can build an operationalization against the same identifiers Eticas uses, which is what makes the scaffold buildable-upon rather than merely readable.

Finally, audit findings are authored once, in a canonical schema, as a single source of truth, and multiple outputs are rendered from that one record for different audiences: a public-facing summary, client reports and regulatory submissions, each projecting the same findings without re-authoring them. Because every finding is tagged to a taxonomy concept, the same record can be re-rendered for a new audience years later in the taxonomy\textquotesingle s stable terms.

\subsection{Why This Is Infrastructure for Others, Not Just for Eticas}\label{why-this-is-infrastructure-for-others-not-just-for-eticas}

Three features distinguish the taxonomy as an infrastructure that others can build on, rather than a self-contained artifact. The foundation layer is a coherent, framework-aligned, operationally scoped risk vocabulary with a clean separation between risk concepts and their mechanisms, published under CC BY 4.0 with stable identifiers, machine-readable distributions, and worked examples. Operationalization attaches the concept identifier and the mechanism identifier, meaning that audits conducted by different practitioners against the same taxonomy remain comparable at the conceptual level regardless of differences in test design. This comparability at the conceptual level is the precondition for cross-provider audit comparability that the field currently lacks. The underlying pattern is also reproducible: the four-layer architecture, the measurement-to-grade chain, the operationalized-entry structure with explicit thresholds, and the visible-gaps principle are specified in sufficient detail to support independent reimplementation. The boundary between the open conceptual layer and the proprietary methodological layer runs along a clearly documented seam: the concept identifiers and the worked examples, mechanism identifiers included, sit on the open side; the mechanism layer across the rest of the catalog, and every calibration, sit on the practitioner side.

The remainder of the paper describes the taxonomy that makes this method scale: its design principles (Section 3), its categories and their public mappings (Section 4), and its alignment to 18 external frameworks (Section 5); it then situates the contribution in the landscape of existing taxonomies and frameworks (Section 6) before concluding.

\section{Design Principles}\label{design-principles}

The Eticas AI Risk Taxonomy has been built from operational experience conducting AI audits across sectors, jurisdictions, and system types. To date this experience spans over fifty AI systems across twelve sectors, among them healthcare, financial services, public administration, law enforcement, human resources, education, energy, and agriculture, and across multiple jurisdictions. The systems assessed cover the technical classes the taxonomy is built to serve: automated decision-making and risk-scoring models (for example clinical-deterioration prediction, recidivism and violence risk assessment, and housing-priority scoring), large language model and retrieval-augmented systems (citizen-facing government chatbots, advisory assistants), computer-vision and biometric systems (facial recognition, diagnostic imaging), and emerging agentic deployments (agentic retrieval assistants, conversational talent-assessment platforms). This range is not incidental to the taxonomy\textquotesingle s shape. The categories and subcategories that are most developed are the ones recurring engagements forced us to make testable, and the sectors we have seen most often are where the mechanism layer runs deepest; the structure is a residue of practice, not a neutral partition of the risk space. The list is illustrative of that practice rather than exhaustive. Its design reflects practitioner constraints: audit teams need navigable structure, test plans need bounded scope per assessment area, compliance reporting needs traceability to obligations, and cross-organizational work needs stable references. Seven principles govern the structure.

\subsection{Three-Level Hierarchy}\label{three-level-hierarchy}

The taxonomy is organized on three levels matched to workflow decisions. Top-level categories (10) provide coarse scoping: a resume-screening tool calls for Bias and Fairness, Privacy and Confidentiality, Governance, and Transparency and Explainability; an autonomous vehicle adds Reliability, Security and Misuse, and Agentic AI. Sub-groups (21) decompose categories into coherent assessment areas with distinct methodologies; under Bias and Fairness, Outcome Disparities needs statistical outcome analysis, Representational Harm needs content analysis, and Dynamic and Systemic Bias needs longitudinal monitoring. Subcategories (70 active) provide operational test scope: each is a bounded risk concept suitable for a dedicated protocol.

The number of subcategories follows from a rule rather than a target. Subcategories are defined at the level where a single test protocol applies: where assessing a risk would require two different methods, we split it; where two risks would always be tested together, we merge them. The current set of 70 active subcategories is the result of applying that rule, and it moves as practice and the technology evolve.

The public surface renders the 8 established categories, all 32 established subcategories, and 17 of the 21 sub-groups --- the four held back being those that are themselves emerging or sit under the emerging Autonomy and Agency category. Emerging subcategories likewise develop in the canonical source until promotion, and the mechanism layer, which is where operational test design attaches, is demonstrated publicly through worked examples rather than published in full. Section 2 explained why this open-core boundary falls where it does.

The 10-21-70 granularity is a deliberate choice. The five OECD principles are too coarse for planning. The NIST AI RMF's many subcategories describe activities rather than risks. The MIT Repository's hundreds of entries overwhelm scoping. AIR 2024's 314 types impose classification overhead. The Eticas structure sits where audit teams can hold the whole space in mind while retaining test-design specificity.

\subsection{Risks Versus Mechanisms: the Contract Surface}\label{risks-versus-mechanisms-the-contract-surface}

This is the most important design decision, and in this version it carries more weight than tidiness, because it is the seam along which the open taxonomy connects to operationalization.

Most taxonomies conflate three ontological categories: risks (abstract harms, such as members of a protected group receiving systematically worse outcomes), mechanisms (the concrete ways a risk surfaces, such as accessibility barriers or quality-of-service disparities), and causes or interventions (unrepresentative training data, parity constraints). Conflation is pervasive: catalogs list ``dataset bias'' (a mechanism) next to ``discriminatory outcomes'' (a harm) next to ``lack of team diversity'' (a root cause). For test design this is debilitating, because an evaluator must specify what to measure (outcomes), how to detect it (mechanisms to probe), and what to recommend (causes to address), and a flat list provides no guidance on which is which.

The Eticas taxonomy keeps subcategories as abstract risks and uses a mechanisms field to enumerate manifestations. The subcategory disparate-impact-protected-groups, for instance, defines the harm; its mechanisms field lists intersectional unfairness, accessibility barriers, quality-of-service disparity, and unequal allocation of opportunity. These are not separate risks. They are the distinct surfaces along which the one risk becomes observable.

The payoff is operationalization. The taxonomy defines the conceptual space (the risk and its mechanisms) and publishes the risks; for its worked examples it also sets out the mechanisms by which each risk surfaces --- the narrative enumeration of its manifestations, distinct from how they are tested. The operationalization of those mechanisms lives in the methodology, which declares, per mechanism, what to test and how. The mechanisms field is the join. Section 2 made this concrete: the same risk - assessed as pii-leakage in the validation run against taxonomy v2.0.0, carried in the current taxonomy as sensitive-information-leakage - counts four mechanisms in v3.0.0 and is operationalized along two of them (disclosure and memorization) with different probes and, importantly, different severity calibrations. The ontological discipline is what makes that clean handoff possible; without it, every operationalization re-litigates what the risk even is.

This mirrors mature ontologies elsewhere: medicine separates diseases from symptoms, signs, and treatments; software engineering separates defects from failure modes, symptoms, and fixes. This taxonomy applies the same discipline to AI risk and demonstrates the seam through its worked examples so others can build on it.

\subsection{Maturity Levels and Publication Strategy}\label{maturity-levels-and-publication-strategy}

Each concept carries a maturity annotation, established or emerging. Established concepts (stable definitions, framework mappings, a populated mechanisms field) form the public scaffold under CC BY 4.0, down to the subcategory level; emerging concepts develop in the canonical source until they meet that bar. The mechanism layer stays internal at every maturity --- both the enumeration of how a risk surfaces and its operationalization (the probes, the kinds of checks, the metric definitions and severity bands) --- and is shown publicly only through the worked examples. The v3.0.0 counts make the split precise: of the 10 active categories, 8 are established and 2 emerging; of the 21 active sub-groups, 17 are rendered on the public surface; of the 70 active subcategories, 32 are established and 38 emerging. A further set of retired concepts (one category, seven sub-groups, twenty-five subcategories) is preserved in the canonical source for institutional memory and excluded from the active counts, with retired identifiers kept resolving as redirects. This staged promotion prevents premature publication while letting the taxonomy evolve as a living resource.

Agentic AI illustrates the path: it has been added at v1.1 as emerging with two sub-groups and seven subcategories, internal-only; promoted at v1.2 to established and public after framework mappings and pilot operationalization accumulated; then surfaced as the eighth established category. Two categories remain emerging: Autonomy and Agency (human-AI relationship risks: overreliance, automation bias, loss of meaningful human control) and Organisational Readiness (organizational capacity, skills, and leadership for AI deployment). The public About page documents that the full conceptual structure spans 10 categories while the public surface shows eight, framing this as deliberate strategy rather than incomplete coverage.

Because the taxonomy reflects our own audit experience, its boundaries are shaped by the sectors and systems we have worked on, and we expect them to change as others apply it. This is why the conceptual layer is open and versioned: revision through use is the intended mechanism, not a sign that the structure is unfinished.

\subsection{Lifecycle Stage Annotation}\label{lifecycle-stage-annotation}

Each subcategory is tagged by where in the pipeline the risk surfaces or is best assessed: pre-processing (data collection and curation), in-processing (training, architecture, tuning), or post-processing (deployment, inference, monitoring). This enables workflow filtering: a pre-deployment review filters to what is assessable from available artifacts, while ongoing monitoring queries post-deployment subcategories. The tagging encodes the practitioner reality that assessment opportunities depend on development stage, and access to data.

\subsection{Audit-Oriented Granularity}\label{audit-oriented-granularity}

The structure emerged from observed audit planning load. Comprehensive system assessments typically engage five to twelve categories, two to four sub-groups within each, producing scopes of roughly ten to thirty subcategories per engagement. Each subcategory is designed to be assessable through a single coherent protocol without recursive decomposition; if assessing it would require two different methodologies in sequence, it is split. This is what distinguishes a practitioner taxonomy from an academic catalog: navigability and bounded scope are the design targets, not exhaustive coverage.

\subsection{Three-Tier Framework Mapping}\label{three-tier-framework-mapping}

The taxonomy maps to 18 frameworks in three tiers carrying explicit epistemic weight. The compliance tier (EU AI Act, ISO/IEC 42001:2023, AIUC-1, CoE Convention, and IEEE 7001, 7002, and 7003) provides regulatory justification, grouping legally binding regulations with directly conformity-assessable standards. The reference tier (NIST AI RMF, NIST 600-1, OECD, TC260, CSA Agentic Profile, OWASP, and IEEE 2894) demonstrates alignment with established practice and soft-law instruments. The academic and vocabulary tier (MIT V4, AIR 2024, IBM Atlas, W3C DPV) provides conceptual grounding and research interoperability. A mapping to EU AI Act Article 10 carries different weight from a mapping to a research subdomain, and the tiering makes that explicit. Mappings use SKOS match-type qualifiers (exactMatch, closeMatch, broadMatch, narrowMatch, relatedMatch); the public surface foregrounds the strongest matches while the canonical source retains full detail and per-mapping citations. Framework alignment is maintained through manual per-category review and is treated as tentative and subject to revision as the external frameworks themselves evolve; a partially automated periodic re-alignment process is under consideration.

\subsection{Open Semantic Infrastructure}\label{open-semantic-infrastructure}

The taxonomy follows W3C best practice. Each concept has a stable URI under \url{https://taxonomy.eticas.ai/risk/}, committed to stability from v1.0 so that citations and tools do not break across versions. The canonical format is SKOS, expressing hierarchy, labels, definitions, and mappings in machine-readable form. The public distribution provides the established top-level categories and subcategories, as well as the sub-groups, as SKOS/Turtle and JSON-LD at the /dist/ path, under CC BY 4.0 with a machine-readable license declaration; per-concept landing pages render the same content humans and machines consume. The emerging concepts and the full mechanism layer live in an internal distribution (/risk-internal/). This is the open-core boundary: the conceptual scaffold, down to established subcategories, is open and reusable, and the methodology layer (Section 2) attaches to it through stable join points (the open concept identifiers and the mechanism identifiers, disclosed in full for the worked examples) rather than private convention.

\section{The Taxonomy: Categories and Their Public Mappings}\label{the-taxonomy-categories-and-their-public-mappings}

This section walks through the eight established, publicly surfaced categories, with definitions, sub-groups, strongest framework anchors, and what makes each distinctive, then notes the two emerging categories. We use disparate-impact-protected-groups as the worked example for subcategory structure; it is one of the subcategories we publish in full as an illustration.

\subsection{Bias and Fairness}\label{bias-and-fairness}

\textbf{Definition}. Risks that AI systems produce or perpetuate unfair, discriminatory, or unequal outcomes, treatment, or representation across individuals or groups, particularly those defined by protected characteristics.

\textbf{Sub-groups}. Outcome Disparities (systematically different outcomes, service quality, or allocation across groups); Representational Harm (stereotyping, demeaning portrayals, erasure); Dynamic and Systemic Bias (feedback loops, drift, systemic reinforcement).

\textbf{Strongest framework anchors}. EU AI Act Article 10 (data governance addressing bias in high-risk systems); NIST AI RMF harmful-bias characteristic; NIST 600-1 category 6 (harmful bias and homogenization); ISO/IEC 42001 Annex A.5 (impact assessment); W3C DPV Discrimination; MIT V4 discrimination subdomains; AIR 2024 societal-risk categories; IBM Atlas fairness dimension; IEEE Std 7003-2024 (algorithmic bias considerations, category level). (Note that EU AI Act Article 5(1)(c), the prohibition on social scoring, is a distinct and narrower anchor; it is not the general high-risk bias obligation, which is Article 10.)

\textbf{Distinctiveness}. The three-way decomposition tracks real methodological boundaries: outcome disparities need quantitative analysis with demographic data; representational harm needs qualitative content analysis and community consultation; dynamic bias needs longitudinal monitoring. Folding these into one ``fairness'' bucket obscures that these are different audits.

\textbf{Worked example}: disparate-impact-protected-groups. This subcategory is the template for understanding subcategory structure, and we publish it in full.

\begin{itemize}
\tightlist
\item
  URI: \url{https://taxonomy.eticas.ai/risk/disparate-impact-protected-groups}
\item
  Broader concept: bias-outcome-disparities (sub-group)
\item
  Maturity: established
\item
  Definition: the system produces systematically different (typically worse) outcomes for individuals belonging to protected groups, including across multiple intersecting attributes (race, gender, age, disability, and others). Encompasses unequal allocation of opportunity, quality-of-service disparities, intersectional unfairness, and accessibility barriers.
\item
  Scope: ALL system types.
\item
  Lifecycle stage: post-processing (single stage; assessable in deployment and monitoring).
\item
  Mechanisms (in source order): intersectional-unfairness (compounded disadvantage at the intersection of protected characteristics); accessibility-barriers (design that excludes people with disabilities); quality-of-service-disparity (degraded performance for some groups); allocation-of-opportunity (uneven distribution of benefits, jobs, credit, services).
\item
  Mappings (ten entries across eight frameworks): MIT subdomain-1.1 (closeMatch) and subdomain-1.3 (closeMatch); NIST 600-1 harmful-bias (broadMatch); NIST AI RMF harmful-bias (closeMatch); EU AI Act Article 5(1)(c) (closeMatch) and Article 10 (closeMatch); W3C DPV Discrimination (closeMatch); ISO/IEC 42001 A.5.2 (closeMatch); AIR 2024 discrimination-and-bias, protected (closeMatch); IBM Atlas output-fairness (closeMatch).
\end{itemize}

Three points about this entry. First, the mappings are almost entirely closeMatch, with one broadMatch (NIST 600-1, whose harmful-bias-and-homogenization category is broader than this single subcategory) and no exactMatch anywhere. The taxonomy deliberately avoids asserting strict equivalence between a risk concept and, for example, a process control such as ISO A.5.2 (``conduct an impact assessment''), because a harm and an activity are not the same kind of thing. The conservative coding is a credibility choice and it is the source of record. Second, MIT and the EU AI Act each carry two mappings, because a single Eticas subcategory can legitimately correspond to more than one concept in a larger framework; the crosswalk records all of them. Third, the mechanisms field is doing the work flagged in Section 3.2: it enumerates the conceptual space (four manifestations) that an operationalization layer later binds tests to. Section 2 took the Privacy counterpart, PII leakage, all the way through that binding; this subcategory is itself one of the operationalized bias-and-fairness entries and follows the identical pattern. (As Section 3.6 notes, these mappings are maintained by manual review by Eticas auditors, and are revised as the external frameworks change; they should be read as current best alignment, not fixed equivalence.)

\subsection{Privacy and Confidentiality}\label{privacy-and-confidentiality}

Definition. Risks that AI systems collect, process, expose, or leak personal or confidential information in violation of privacy rights, data-protection law, or confidentiality clauses and commitments.

Sub-groups. Data Collection and Use (excessive collection, purpose-limitation violations, secondary use); Data Protection (unauthorized access, weak security, re-identification); Model-level Privacy (training-data extraction or memorization, model inversion, re-identification from outputs).

Strongest framework anchors. AIUC-1 data-and-privacy domain; multiple W3C DPV privacy concepts; NIST 600-1 category 4 (data privacy); EU AI Act Article 10(5) (personal data in high-risk systems); ISO/IEC 42001 Annex A.7 (data); OECD privacy principle; TC260 privacy provisions; IEEE Std 7002-2022 (data privacy process, category level).

Distinctiveness. Privacy is a regulatory convergence point, and the structure reflects it. Model-level Privacy is the AI-specific edge: training-data extraction and model inversion exploit statistical properties of models, exposing information without ever touching a data store, and conventional data-protection controls do not reach them. This is also one of the two dimensions Eticas has fully operationalized (Section 2), where sensitive-information-leakage - the v3.0.0 concept broadened from the pii-leakage subcategory of Section 2's validation to cover personal alongside confidential or proprietary data - is decomposed into four mechanisms in the taxonomy, of which two (disclosure and memorization) are currently operationalized; cross-customer data contamination and system-prompt extraction remain declared but not yet operationalized, the visible empty slots of Section 2.5.

\subsection{Reliability}\label{reliability}

\textbf{Definition}. Risks that AI systems produce incorrect, inconsistent, or unreliable outputs, or fail to maintain acceptable performance under expected conditions and perturbations.

\textbf{Sub-groups}. Output Integrity and Robustness (accuracy, consistency, alignment with expected behavior); Operational Resilience (emerging) (recovery, graceful degradation, maintained function after disruption); Measurement Validity (emerging) (whether what the system measures captures the construct it claims to assess; seeded by the construct-validity subcategory, not yet operationalized).

\textbf{Strongest framework anchors}. EU AI Act Article 15 (accuracy and robustness); ISO/IEC 42001 Annex A.5; NIST AI RMF valid-and-reliable characteristic; NIST 600-1 category 2 (confabulation); OECD robustness principle; AIR 2024 system and operational risks; IBM Atlas reliability dimensions.

\textbf{Distinctiveness}. Separating Operational Resilience (emerging) from Output Integrity and Robustness recognizes that ``performs correctly'' and ``fails safely'' are different audits with different methods. Resilience is fragmented across existing frameworks (NIST folds it into ``secure and resilient''); treating it as a coherent assessment area is a contribution, pending operationalization. Measurement Validity, promoted to a sub-group in v3.0.0, adds a third methodological boundary: whether the measured proxy captures the claimed construct is a distinct question from whether outputs are accurate.

\subsection{Governance}\label{governance}

\textbf{Definition}. Risks from inadequate organizational structures, accountability mechanisms, documentation, or compliance processes for responsible AI development and deployment.

\textbf{Sub-groups}. Accountability (responsibilities, decision rights, redress); Documentation and Auditability (records, technical documentation, audit trails); Compliance and Process (adherence to regulations, standards, internal policy).

\textbf{Strongest framework anchors}. ISO/IEC 42001 Annex A.2 (AI policies) and A.3 (internal organization), the closest of the governance mappings; EU AI Act Articles 9, 11, and 17 (risk management, technical documentation, quality management); NIST AI RMF Govern function; OECD accountability principle; CoE Convention oversight obligations.

\textbf{Distinctiveness}. Governance receives less technical attention than algorithmic risk but is the primary failure mode in real incidents: systems discriminate not because bias is unsolvable but because no process required fairness validation before deployment. Surfacing organizational and process risks as first-class is a stance grounded in audit experience. This is the most regulatorily anchored category, since ISO 42001 certification turns directly on these controls.

\subsection{Security and Misuse}\label{security-and-misuse}

\textbf{Definition}. Risks from malicious attacks exploiting AI vulnerabilities, conventional cybersecurity weaknesses in AI infrastructure, or intentional harmful use of AI capabilities.

\textbf{Sub-groups}. AI-specific Attacks (adversarial examples, data poisoning, model extraction); System Security (access control, infrastructure, conventional cybersecurity in AI contexts); Harmful Misuse (intentional harm, dual-use).

\textbf{Strongest framework anchors}. OWASP Top 10 for Agentic Applications (multiple); NIST 600-1 category 9 (information security); EU AI Act Article 15(1) (cybersecurity); ISO/IEC 42001 with ISO 27001 integration; NIST AI RMF secure-and-resilient; CSA Agentic Profile; TC260 security provisions.

\textbf{Distinctiveness}. The category separates intentional attacks from unintentional failures (which live in Reliability), because threat modeling and quality assurance are different disciplines. Several OWASP agentic categories map here, illustrating how agentic deployment expands the attack surface (tool misuse, memory poisoning, unsafe code execution).

\subsection{Environmental Impact}\label{environmental-impact}

\textbf{Definition}. Risks that AI systems cause environmental harm through energy use, carbon emissions, resource consumption, or electronic waste, especially at training and deployment scale.

\textbf{Sub-groups}. None; flat category at present.

\textbf{Strongest framework anchors}. NIST 600-1 category 5 (environmental impacts); OECD inclusive-growth and sustainability principle; MIT V4 environmental subdomain; IBM Atlas environmental impact. This category shows the cleanest cross-framework convergence in the taxonomy.

\textbf{Distinctiveness}. Environmental risk shares one assessment methodology (lifecycle analysis of consumption and emissions), so the flat structure is appropriate. The category also demonstrates the taxonomy's ability to absorb newly salient risks: environmental concerns were absent from early taxonomies and now command broad consensus.

\subsection{Transparency and Explainability}\label{transparency-and-explainability}

\textbf{Definition}. Risks that AI systems operate without adequate explanation of their functioning, decision rationale, or performance, limiting stakeholder understanding and accountability.

\textbf{Sub-groups}. Explainability (technical interpretability, decision rationale, feature importance); Communication and Disclosure (user-facing disclosure, capability and limitation transparency).

\textbf{Strongest framework anchors}. EU AI Act Article 86 (the right to explanation of individual decisions); EU AI Act Article 13 (transparency and information to deployers); ISO/IEC 42001 Annex A.8 (information for interested parties); NIST AI RMF explainable-and-interpretable; OECD transparency principle; W3C DPV explanation concepts; CoE Convention information obligations; IEEE Std 7001-2021 (transparency of autonomous systems) and IEEE Std 2894-2024 (architectural framework for explainable AI), both at category level.

\textbf{Distinctiveness}. The category separates technical explanation from stakeholder communication because they serve different functions with different methods. The regulatory anchor is precise: EU AI Act Article 86 grants affected persons the right to obtain from the deployer, in the Act's own words, ``clear and meaningful explanations of the role of the AI system in the decision-making procedure and the main elements of the decision taken.'' This is narrower and more operational than the broader, and contested, GDPR debate over a ``right to explanation,'' and the taxonomy maps to the AI Act language specifically. Semantic alignment with W3C DPV explanation concepts lets compliance and audit tooling share concept identifiers.

\subsection{Agentic AI}\label{agentic-ai}

\textbf{Definition}. Risks specific to AI systems that autonomously plan, invoke external tools, execute multi-step action chains, and interact with other agents or environments with reduced human involvement.

\textbf{Sub-groups}. Autonomous Actions and Tool Use (autonomous decisions, tool and API invocation, real-world action, unintended consequences); Multi-agent Integrity (emergent behavior, coordination failure, conflicting objectives, cascading failure across agent networks).

\textbf{Strongest framework anchors}. CSA Agentic Profile (NIST AI RMF v1); OWASP Top 10 for Agentic Applications 2026 (multiple); AIUC-1 agent-specific controls.

\textbf{Distinctiveness}. This is the newest established category and the locus of the governance-gap finding. Agentic systems decompose goals, invoke tools, persist state, adapt plans, and interact with other agents, which creates risk profiles absent from passive prediction systems: an agent with deletion access may over-interpret an instruction; an agent network with conflicting objectives may produce emergent behavior no agent intended; an agent acting at machine speed forecloses meaningful per-decision oversight. Among major audit-oriented taxonomies, Eticas is among the first to treat Agentic AI as a first-class category (the IBM Atlas also includes an agentic category), and the frameworks it maps to all emerged in 2025 and 2026. The thinness of mappings everywhere else is the finding, developed in Section 5.

\subsection{Emerging Categories}\label{emerging-categories}

Two categories exist at emerging maturity, documented internally pending operationalization. Autonomy and Agency address the human-AI relationship (overreliance, automation bias, deskilling, loss of meaningful human control); it is distinct from Agentic AI, which concerns agent-specific technical risk regardless of the human relationship. Organisational Readiness addresses organizational capacity for responsible deployment (skills, leadership, change management). Both have initial mappings and definitions but await further operational validation before public promotion.

\section{Framework Alignment Analysis}\label{framework-alignment-analysis}

This section examines the taxonomy as a Rosetta Stone across the governance landscape: a three-tier crosswalk, a completeness check against NIST 600-1, the structural variations in how frameworks organize risk, and the agentic governance gap.

\subsection{The Three-Tier Crosswalk}\label{the-three-tier-crosswalk}

Table 3 summarizes mapping coverage across the eight public categories. A cell marks whether at least one subcategory maps to the framework at exactMatch or closeMatch (C), only at a weaker broad, narrow, or related match (R), or not at all.

\begin{table}[htbp]
\centering
\includegraphics[width=\textwidth]{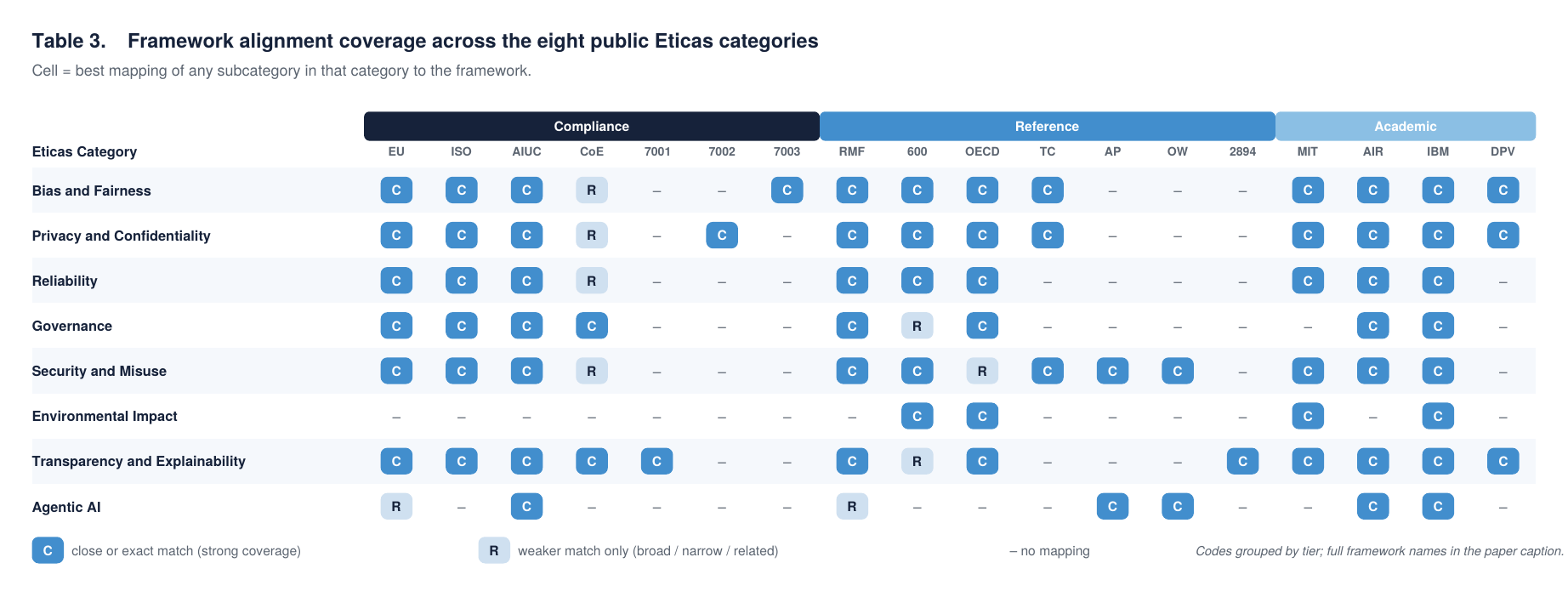}
\caption{Framework alignment coverage across the eight public categories. Coverage is broadest for Bias and Fairness, Privacy and Confidentiality, and Security and Misuse; Agentic AI is covered only by purpose-built frameworks.}
\label{tab:crosswalk}
\end{table}

Framework codes, grouped by tier. Compliance: EU = EU AI Act; ISO = ISO/IEC 42001; AIUC = AIUC-1; CoE = Council of Europe Convention; 7001 = IEEE 7001; 7002 = IEEE 7002; 7003 = IEEE 7003. Reference: RMF = NIST AI RMF; 600 = NIST AI 600-1; OECD = OECD Principles; TC = TC260; AP = CSA Agentic Profile; OW = OWASP Agentic Top 10; 2894 = IEEE 2894. Academic: MIT = MIT Repository V4; AIR = AIR 2024; IBM = IBM Atlas; DPV = W3C DPV.

Bias and Fairness, Privacy and Confidentiality, and Security and Misuse achieve the broadest coverage (close matches in ten or more frameworks), reflecting consensus on these domains. Environmental Impact achieves strong but narrower coverage, concentrated in frameworks that explicitly address sustainability. The CSA Agentic Profile and OWASP concentrate, as expected, in Agentic AI and Security and Misuse, and the four IEEE standards sit at category level on Bias and Fairness (7003), Privacy and Confidentiality (7002), and Transparency and Explainability (7001 and 2894). Agentic AI shows the starkest pattern: strong coverage only in the agentic-specific frameworks, examined in Section 5.4.

\subsection{NIST 600-1 Completeness Check}\label{nist-600-1-completeness-check}

Where Table 3 runs outward, asking which frameworks recognize each Eticas category, Table 4 runs inward: it takes an external framework's complete enumeration and tests Eticas against it. The direction matters, because a coverage map cannot reveal a category Eticas lacks (anything missing simply never appears as a row), whereas a completeness check can. We use NIST 600-1 as the instrument because it is the rare framework suited to the task: a bounded, enumerable set of twelve discrete risk areas, risk-domain organized like the Eticas taxonomy (Section 5.3) so the comparison is like-for-like, recent and generative-AI-specific, and authoritative without being Eticas-aligned, so a clean bill of health would not be self-serving.

\begin{table}[htbp]
\centering
\includegraphics[width=\textwidth]{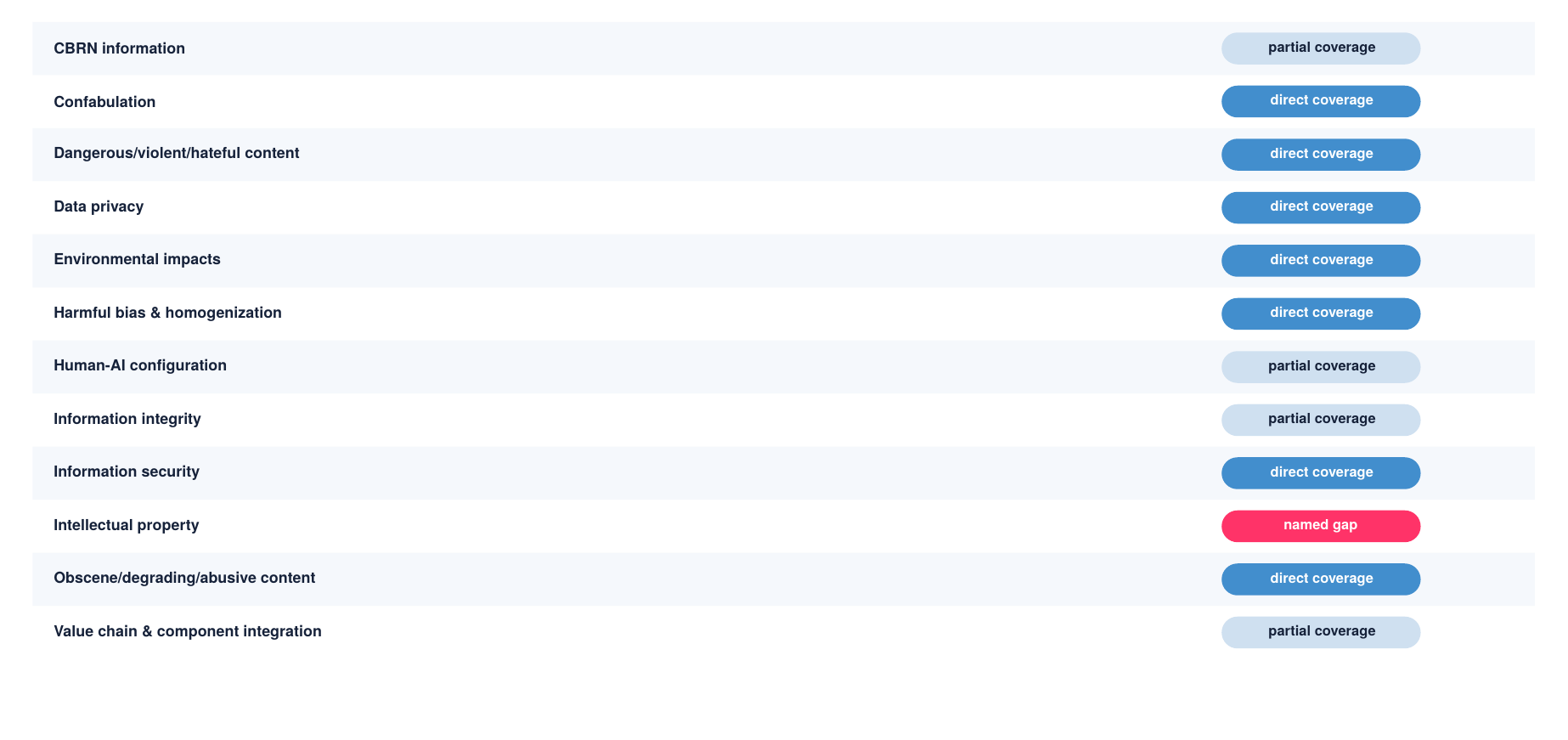}
\caption{NIST AI 600-1 coverage analysis. Seven of the twelve risk areas have direct coverage, four are partial, and intellectual property is a named gap.}
\label{tab:nist}
\end{table}

Seven of twelve areas have direct coverage, four are partial, and one (intellectual property) is a true gap. The audit-tractable portion of that gap, the exposure of confidential or proprietary information through model outputs, is captured in the taxonomy by the established subcategory sensitive-information-leakage, which absorbed the earlier confidential-information-leakage concept and covers confidential and proprietary data alongside personal data, framed as a confidentiality risk rather than a new intellectual-property category; copyright and patent questions remain outside what a system audit can measure. We state this plainly because it reflects a methodological commitment we call ``named gaps over weak matches'': telling a user that intellectual property requires a supplemental framework is more useful than asserting a tenuous mapping that breaks under scrutiny. This same commitment, as Section 2 showed, is built into the operationalization layer, where an unaddressed mechanism appears as a visible empty slot rather than a silent omission.

\subsection{Four Structural Variations}\label{four-structural-variations}

The 18 frameworks organize risk in four distinct ways, and recognizing this is essential for accurate translation. Risk-domain organization (Eticas, MIT V4, NIST 600-1, AIR 2024) organizes by what can go wrong and aligns with audit workflow. Lifecycle-stage organization (IBM Atlas, NIST RMF functions) organizes by where in the pipeline risk surfaces and aligns with development workflow. Source-of-risk organization (TC260) organizes by where risk originates (endogenous, application, derivative). Principles organization (OECD, CoE) organizes by ethical commitments and enables consensus.

The Eticas risk-domain choice aligns with audit workflow but imposes translation work when mapping to differently organized frameworks; the TC260 mapping is held at category level for this reason. The deeper implication is that no single taxonomy serves all uses: research benefits from comprehensive catalogs, development from lifecycle organization, policy from principles, and audit from risk-domain organization at operational granularity. The ecosystem needs multiple frameworks, not convergence on one, which is also why publishing an open, interoperable, operationalizable taxonomy is more valuable than publishing one more closed catalog.

\subsection{The Agentic AI Governance Gap}\label{the-agentic-ai-governance-gap}

Table 5 sets the publication dates of the frameworks against the emergence of agentic AI as a deployment paradigm.

\begin{table}[htbp]
\centering
\includegraphics[width=\textwidth]{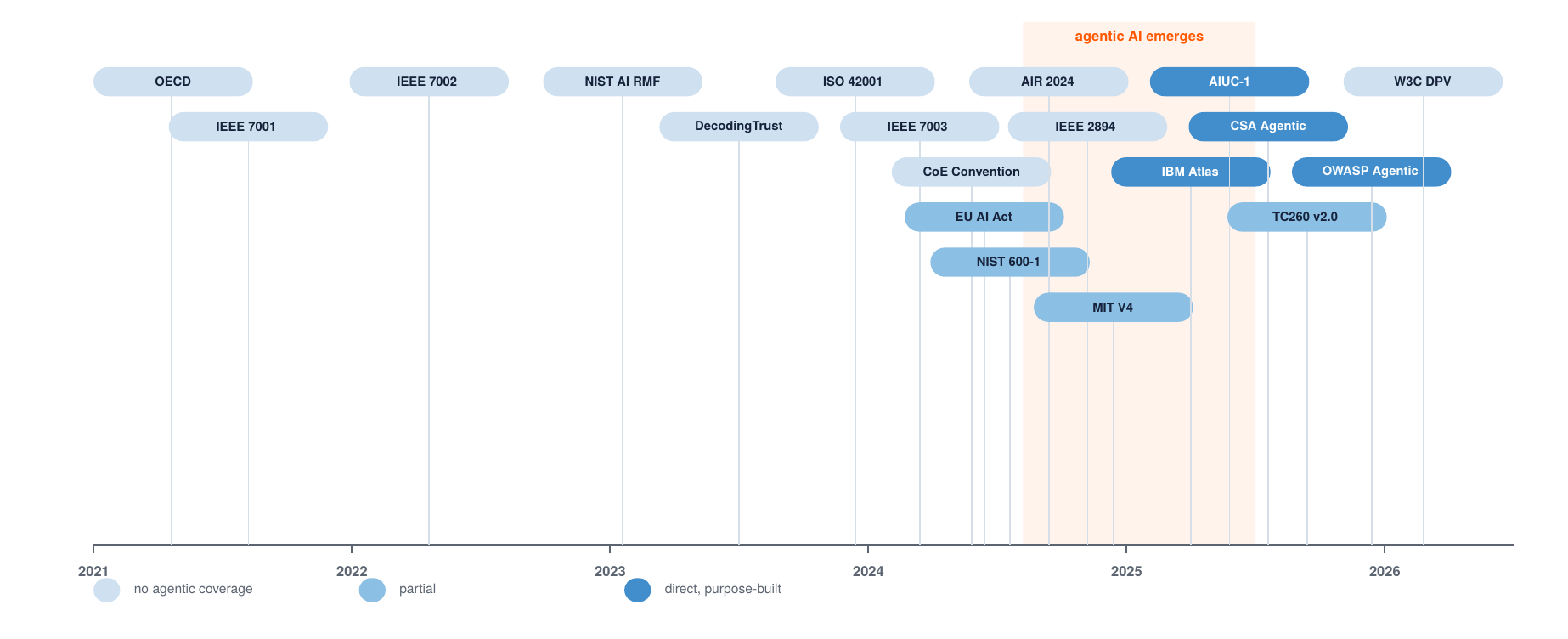}
\caption{Framework publication timeline against agentic AI emergence. Every binding compliance framework predates the emergence band; the only direct-coverage frameworks are purpose-built and date from 2025 onward.}
\label{tab:timeline}
\end{table}

The pattern is stark. Every compliance-tier framework providing binding obligations was finalized before or as agentic AI emerged, and most reference-tier frameworks predate it too. Only the purpose-built frameworks (AIUC-1, the CSA Agentic Profile, OWASP) and the IBM Atlas provide direct coverage, and all date from 2025 or later. This is a structural gap, not an incidental one: governance frameworks emerge on multi-year cycles through regulatory negotiation and standards consensus, while agentic AI emerged as a deployment paradigm in roughly eighteen months.

Treating Agentic AI as first-class makes the gap legible to evaluators: an audit of an agentic system can see, from the mappings alone, that compliance assessment against the EU AI Act or ISO 42001 will be partial and must be supplemented with OWASP, the CSA Agentic Profile, or AIUC-1. The positioning also supports the three response paths open to regulators (amend existing frameworks, issue interpretive guidance, or develop new agent-specific frameworks): the mapping shows precisely where existing frameworks need extension, and the category supplies vocabulary that guidance can reference. Independent analyses confirm the gap, finding the EU AI Act ``not ready for agents'' (Gardhouse and Oueslati, 2026) and documenting compliance difficulties for AI agents under EU law (Nannini et al., 2026).

\section{Related Work}\label{related-work}

AI risk taxonomies now number in the dozens, organized by different principles and serving different constituencies. We structure this review by orientation: research and meta-review taxonomies, vendor and practitioner frameworks, regulatory and standards frameworks, semantic and vocabulary frameworks, and the recent agentic-specific frameworks. Running through the review is a single comparative question that motivates this paper: do existing frameworks demonstrate a path from its risk concepts to executed measurement, or do they stop at the catalog?

\subsection{Research and Meta-Review Taxonomies}\label{research-and-meta-review-taxonomies}

The most comprehensive recent effort is the MIT AI Risk Repository V4 (Slattery et al., 2026), a meta-review cataloguing many hundreds of risks extracted from 74 frameworks spanning government regulations, corporate policies, and academic taxonomies. The Repository structures risks along two axes: a causal taxonomy classifying by entity (human, AI), intent (intentional, unintentional), and timing (pre- or post-deployment); and a domain taxonomy organizing seven societal impact areas into subdomains. Published in Patterns with an interactive web tool under CC BY 4.0, it is rigorous empirical foundation for the field and the natural reference point against which any specialized taxonomy should situate itself.

Its comprehensiveness, however, limits operational use. It preserves the heterogeneity of its sources rather than normalizing to one vocabulary; equivalent concepts from different frameworks appear as separate entries. It mixes abstraction levels, placing mechanisms (``dataset bias''), observable harms (``discriminatory outcomes''), and organizational root causes (``lack of diversity in development teams'') as coordinate entries. And at many hundreds of entries it exceeds the cognitive load of audit scoping. None of this is a flaw in a meta-review; it is the cost of comprehensiveness. But it means the Repository is a map of the territory, not a tool for traversing it. It does not, and does not aim to, show how any one risk is measured.

AIR 2024 (Zeng et al., 2024) takes a bottom-up approach, extracting 314 risk types from eight government regulations and sixteen corporate policies, organized in a four-tier ontology. The accompanying AIR-Bench operationalizes the taxonomy into a safety benchmark aligned to regulatory language, which is a genuine and unusual step toward measurement. AIR's strength is traceability to compliance language; its costs are granularity (314 types impose classification overhead) and the gap between policy-derived phrasing and test design. AIR-Bench shows that a taxonomy can drive a benchmark, but it targets model-level safety evaluation rather than the broader, system-and-context audits that regulatory compliance requires.

\subsection{Vendor and Practitioner Frameworks}\label{vendor-and-practitioner-frameworks}

IBM AI Risk Atlas (Bagehorn et al., 2025) organizes risks by lifecycle stage (input, inference, output, non-technical, and agentic) and integrates with Risk Atlas Nexus, an open-source tooling suite of ontologies, knowledge graphs, and AI-assisted compliance workflows. This is among the most operationally serious of the published frameworks, and notably it already includes an agentic category. Its lifecycle organization aligns with development workflows but complicates compliance crosswalks, since regulators organize by domain of harm. Its emphasis on automated detection underweights the governance and process controls where human judgment remains essential and where, in our audit experience, many real failures actually originate.
The Eticas taxonomy presented in this paper has itself been integrated into Risk Atlas Nexus as one of its external taxonomy resources.

Beyond IBM, most practitioner methodologies remain unpublished. Specialist audit firms hold proprietary frameworks and do not expose either their risk vocabularies or, more importantly, their operationalization. The broader accountability literature has mapped this ecosystem and its incentives (Raji et al., 2020; Mökander and Floridi, 2021; Costanza-Chock et al., 2022), and artifact standards such as datasheets and model cards (Gebru et al., 2021; Mitchell et al., 2019) structure documentation, but neither supplies an operational risk vocabulary. The field therefore lacks an open practitioner taxonomy at audit-suitable granularity that also shows its operationalization. That absence is one of the gaps this paper addresses.

\subsection{Regulatory and Standards Frameworks}\label{regulatory-and-standards-frameworks}

The EU AI Act (Regulation 2024/1689) establishes risk-based regulation with prohibited practices (Article 5), high-risk categories (Annex III), and essential requirements distributed across Articles 9 to 15. Article 86 establishes a right to explanation for certain decisions by high-risk systems. As a regulation, it provides binding obligations but not an operational taxonomy: Article 15 bundles ``accuracy, robustness and cybersecurity,'' and the Act specifies that systems achieve ``appropriate'' levels without defining an assessment methodology. Harmonized standards under CEN/CENELEC JTC 21 will supply technical specifications, but they will focus on conformity assessment procedures rather than risk vocabulary. The Act's value to taxonomy work lies in compliance anchors: mappings to specific Articles provide regulatory justification for including assessment criteria.

ISO/IEC 42001:2023, the first international AI management system standard, uses Plan-Do-Check-Act methodology and enumerates Annex A controls across policies (A.2), internal organization (A.3), resources (A.4), impact assessment (A.5), data (A.7), and information for interested parties (A.8). As a certifiable standard it drives audit demand. But it operates at the level of the organization's management processes, not the AI system's measured behavior: it governs whether an organization has a policy or an impact-assessment procedure, not what a model does when tested. Its control-based structure therefore does not map directly to a risk taxonomy: A.5 requires impact assessment without enumerating risk categories, so evaluators need a supplemental vocabulary bridging from organizational controls to testable risk domains.

NIST AI 600-1 (Autio et al., 2024) defines 12 risk categories for generative AI: CBRN information, confabulation, dangerous or violent or hateful content, data privacy, environmental impacts, harmful bias and homogenization, human-AI configuration, information integrity, information security, intellectual property, obscene or degrading or abusive content, and value chain and component integration. Each carries suggested actions mapped to the AI RMF functions. Its 12-category scope is manageable but bounded to generative systems, and as Section 5 showed, it makes a useful instrument for a completeness check against the Eticas taxonomy.

The earlier NIST AI RMF (Tabassi, 2023) organizes around seven trustworthiness characteristics and four functions (Govern, Map, Measure, Manage). Two features limit its use as a risk taxonomy. First, it is process orientation rather than risk enumeration: it guides how to approach risk management without specifying which risks to manage, and its characteristics overlap by design. Second, its four functions describe activities an organization performs rather than properties a system exhibits, so its unit of analysis is the organization's risk-management process, not the model's measured behavior. This is the same level-of-operation gap found in the management-system standards: the framework governs whether risk management happens, not what testing the system reveals. For audit planning that needs clear boundaries between assessment areas, and findings expressed as measured system properties, it requires supplementation with a risk-domain taxonomy.

The OECD AI Principles (OECD/LEGAL/0449, updated May 2024) provide five value-based principles with broad international consensus, and the Council of Europe Framework Convention (CETS 225, 2024) is the first binding international AI treaty. Both are normative foundations at an abstraction level that prevents direct operationalization; they anchor values rather than tests. TC260 v2.0 (September 2025) organizes risks by source (endogenous, application, derivative) rather than domain of harm, reflecting a different governance philosophy and imposing translation work for cross-jurisdictional practitioners. We map TC260 at category level for the four Eticas categories where clear equivalents exist. The taxonomy also maps four IEEE standards: the conformity-assessable IEEE Std 7001-2021 (transparency of autonomous systems), 7002-2022 (data privacy process), and 7003-2024 (algorithmic bias considerations), placed in the compliance tier, and the advisory IEEE Std 2894-2024 (architectural framework for explainable AI), placed in the reference tier. Each is mapped at category level on its relevant domain.

\subsection{Semantic and Vocabulary Frameworks}\label{semantic-and-vocabulary-frameworks}

The W3C Data Privacy Vocabulary v2.3 with AI Extension (Pandit and Golpayegani, 2026) is the most mature example of AI risk concepts published as semantic web infrastructure, using SKOS with stable URIs to represent GDPR and AI Act compliance concepts. DPV demonstrates exactly what semantic infrastructure enables: programmatic consumption, formal reasoning, and machine-discoverable compliance knowledge. Its scope emphasizes legal compliance concepts (obligations, roles, purposes) rather than technical risk assessment at evaluation granularity. The Eticas taxonomy deliberately adopts DPV's infrastructure pattern (SKOS, stable URIs, open licensing) while targeting audit operationalization.

\subsection{Agentic-Specific Frameworks (2025 to 2026)}\label{agentic-specific-frameworks-2025-to-2026}

The emergence of agentic AI from late 2024 produced a rapid standardization response. AIUC-1 (2025) positions itself as a `SOC 2 for AI agents,' with controls across six domains and a certification model including independent auditing and technical testing, mapped to the EU AI Act, NIST AI RMF, ISO 42001, MITRE ATLAS, and OWASP. As a SOC 2-style certification its primary unit of analysis is the organization's controls (whether safeguards exist and are operating), the same level-of-operation orientation seen in the management-system standards, even though its certification model reaches further toward the system than ISO 42001 by incorporating independent technical testing. That technical-testing component is what an audit measures on the system; the surrounding controls are what it verifies about the organization. For a risk-domain vocabulary that expresses findings as measured system properties, the control set still needs a supplemental mapping. The Cloud Security Alliance published an Agentic Profile extending NIST AI RMF 1.0 for agentic systems (autonomy-tier classification, tool-use risk, runtime behavioural governance, delegation-chain accountability), organized around the Govern-Map-Measure-Manage structure and aligned with the CSA AI Controls Matrix. The OWASP Top 10 for Agentic Applications 2026 (December 2025), developed by a large practitioner community with expert review, enumerates ten agent-specific risk categories, from agent goal hijacking and tool misuse to cascading planning failures, memory poisoning, and insufficient observability.

These frameworks demonstrate responsive standardization, and their value is real, but their recency is the point: the compliance backbone predates them by one to three years. Section 5 developed this as a structural finding.

\subsection{Positioning the Eticas Taxonomy}\label{positioning-the-eticas-taxonomy}

As the above clearly shows, the landscape leaves a specific opening. On comprehensiveness versus usability, meta-reviews and large inventories serve discovery but exceed audit cognitive load. Principles frameworks enable consensus but lack operational specificity. The Eticas 10-21-70 structure is a deliberate middle path. On ontological clarity, most taxonomies conflate risks with mechanisms and causes, whereas Eticas separates them and uses the mechanisms field as an operationalization contract. On semantic infrastructure, few practitioner taxonomies reach DPV's level of SKOS, stable URIs, and open licensing; Eticas adopts it as a baseline. On framework alignment, the three-tier mapping to 18 frameworks provides a Rosetta Stone across compliance, reference, and academic vocabularies. On unit of analysis, several of the most influential compliance instruments (the management-system standard ISO/IEC 42001, the process-oriented NIST AI RMF, and the SOC 2-style certification AIUC-1) operate primarily at the level of the deploying organization's controls and processes, governing whether risk management happens rather than what testing reveals about the model.

Two frameworks go furthest toward measurement, AIR 2024\textquotesingle s AIR-Bench and IBM\textquotesingle s Risk Atlas, but even they stop at model-safety benchmarking, not system-and-context audits; and automated detection, not graded findings. The Eticas taxonomy is built to express findings as measured properties of the system, which is the level at which an audit produces evidence. And on the decisive axis, operationalization, the Eticas taxonomy is, to our knowledge, the first openly published audit taxonomy accompanied by a documented, benchmark-validated path from its concepts to graded audit findings. The preceding sections substantiate these claims, with Section 2 carrying the operationalization argument.

\section{Conclusion}\label{conclusion}

AI auditing has reached the point where infrastructure matters as much as method. Individual audits can succeed through skilled practitioners applying ad hoc frameworks, but the field cannot accumulate learning, regulators cannot evaluate consistency, and clients cannot compare services without shared, open, and operable foundations. The proliferation of risk taxonomies reflects the demand for shared vocabulary; their fragmentation, and their overwhelmingly theoretical character, reflects the absence of infrastructure built for and validated against the realities of audit practice.

The Eticas AI Risk Taxonomy v3.0.0 is a contribution toward closing that gap. It is practitioner-built and audit-oriented; organized at a granularity tuned to real planning; mapped to 18 external frameworks across compliance, reference, and academic tiers; and published as open semantic infrastructure under CC BY 4.0, with its conceptual scaffold and worked subcategory examples carrying stable identifiers and SKOS and JSON-LD distributions. Its ontological separation of risks from mechanisms is not only a conceptual tidiness but the stable seam along which an operationalization layer attaches, disclosed in full for the worked examples, and we have shown that layer in operation: a four-layer methodology, a codified measurement-to-grade chain, operationalized entries that bind tool-agnostic probes to external benchmarks with disclosed severity thresholds, and an end-to-end validation against the public DecodingTrust benchmark in two dimensions, shown end to end for PII leakage in Section 2. Treating Agentic AI as a first-class category surfaces a structural governance gap, where the major frameworks predate the deployment paradigm they are now asked to govern.

This taxonomy is a work-in-progress, and so we invite the algorithmic-auditing community, regulators, researchers, and AI developers to use it, map their frameworks against it, build their own operationalizations on its open concepts, and propose extensions and corrections. The infrastructure value emerges through use, and the seam is open for building.

The taxonomy is available at \url{https://taxonomy.eticas.ai/}, with purpose, scope, and licence boundary documented at \url{https://taxonomy.eticas.ai/about/} and SKOS distributions at \url{https://taxonomy.eticas.ai/dist/taxonomy.ttl} and \url{https://taxonomy.eticas.ai/dist/taxonomy.jsonld.} Suggested citation: Eticas. (2026). Eticas AI Risk Taxonomy, v3.0.0. \url{https://taxonomy.eticas.ai/risk/.}

\section*{Competing Interests}\label{competing-interests}

The authors are affiliated with Eticas, which provides commercial AI audit services and maintains the proprietary methodology layer and the full subcategory set described in this paper. The conceptual scaffold of the taxonomy is released under CC BY 4.0. This work should be read with that commercial interest disclosed.

\section*{Data and Code Availability}\label{data-and-code-availability}

The public taxonomy, comprising the established top-level categories and subcategories, as well as the sub-groups, with per-concept pages and machine-readable SKOS/Turtle and JSON-LD distributions, is available at \url{https://taxonomy.eticas.ai/} under CC BY 4.0 and currently serves v3.0.0. High-resolution versions of all figures in this paper are available alongside the taxonomy at \url{https://taxonomy.eticas.ai/.} All 32 established subcategories are browsable at their concept URIs, including disparate-impact-protected-groups and sensitive-information-leakage (the v3.0.0 successor of the pii-leakage concept against which the Section 2 validation was run; the retired identifier resolves as a redirect); the operational layer behind the worked examples is disclosed in this paper rather than rendered on the public pages. The emerging subcategories and the full mechanism layer are maintained in an access-gated internal distribution and are not publicly browsable. The DecodingTrust benchmark used for validation is publicly available (Wang et al., 2023). The audit methodology repository and engagement-specific audit data are proprietary and not publicly available; the architecture and operationalization patterns are described in Section 2 in sufficient detail to be independently re-implemented against the public taxonomy.

\clearpage
\section*{References}
\refentry{Artificial Intelligence Underwriting Company (AIUC). (2025). AIUC-1: The AI agent standard. \url{https://www.aiuc-1.com/}}

\refentry{Autio, C., Schwartz, R., Dunietz, J., Jain, S., Stanley, M., Tabassi, E., Hall, P., \& Roberts, K. (2024). Artificial Intelligence Risk Management Framework: Generative Artificial Intelligence Profile (NIST AI 600-1). National Institute of Standards and Technology. \url{https://doi.org/10.6028/NIST.AI.600-1}}

\refentry{Bagehorn, F., Brimijoin, K., Daly, E. M., He, J., Hind, M., Garces-Erice, L., Giblin, C., Giurgiu, I., Martino, J., Nair, R., Piorkowski, D., Rawat, A., Richards, J., Rooney, S., Salwala, D., Tirupathi, S., Urbanetz, P., Varshney, K. R., Vejsbjerg, I., \& Wolf-Bauwens, M. L. (2025). AI Risk Atlas: Taxonomy and Tooling for Navigating AI Risks and Resources. arXiv preprint arXiv:2503.05780. \url{https://arxiv.org/abs/2503.05780}}

\refentry{Bender, E. M., Gebru, T., McMillan-Major, A., \& Shmitchell, S. (2021). On the Dangers of Stochastic Parrots: Can Language Models Be Too Big? Proceedings of the 2021 ACM Conference on Fairness, Accountability, and Transparency (FAccT '21), 610-623. \url{https://doi.org/10.1145/3442188.3445922}}

\refentry{Cloud Security Alliance. (2025). Agentic Profile: A NIST AI RMF Profile for Agentic AI. Cloud Security Alliance. \url{https://labs.cloudsecurityalliance.org/agentic/agentic-nist-ai-rmf-profile-v1/}}

\refentry{Costanza-Chock, S., Raji, I. D., \& Buolamwini, J. (2022). Who Audits the Auditors? Recommendations from a field scan of the algorithmic auditing ecosystem. Proceedings of the 2022 ACM Conference on Fairness, Accountability, and Transparency (FAccT '22), 1571-1583. \url{https://doi.org/10.1145/3531146.3533213}}

\refentry{Council of Europe. (2024). Framework Convention on Artificial Intelligence and Human Rights, Democracy and the Rule of Law (CETS No.~225). Adopted 17 May 2024.}

\refentry{Eticas. (2026). Eticas AI Risk Taxonomy, v3.0.0. \url{https://taxonomy.eticas.ai/risk/}}

\refentry{European Union. (2024). Regulation (EU) 2024/1689 of the European Parliament and of the Council of 13 June 2024 laying down harmonised rules on artificial intelligence (Artificial Intelligence Act). Official Journal of the European Union, L 2024/1689. \url{https://eur-lex.europa.eu/eli/reg/2024/1689/oj/eng}}

\refentry{Gardhouse, K., \& Oueslati, A. (2026, May 5). The EU AI Act is Not Ready for Agents. Tech Policy Press. \url{https://www.techpolicy.press/the-eu-ai-act-is-not-ready-for-agents/}}

\refentry{Gebru, T., Morgenstern, J., Vecchione, B., Vaughan, J. W., Wallach, H., Daumé III, H., \& Crawford, K. (2021). Datasheets for Datasets. Communications of the ACM, 64(12), 86-92. \url{https://doi.org/10.1145/3458723}}

\refentry{Institute of Electrical and Electronics Engineers (IEEE). (2021). IEEE Std 7001-2021: IEEE Standard for Transparency of Autonomous Systems. IEEE. \url{https://standards.ieee.org/ieee/7001/6929/}}

\refentry{Institute of Electrical and Electronics Engineers (IEEE). (2022). IEEE Std 7002-2022: IEEE Standard for Data Privacy Process. IEEE. \url{https://standards.ieee.org/ieee/7002/6898/}}

\refentry{Institute of Electrical and Electronics Engineers (IEEE). (2024a). IEEE Std 7003-2024: IEEE Standard for Algorithmic Bias Considerations. IEEE. \url{https://standards.ieee.org/ieee/7003/7194/}}

\refentry{Institute of Electrical and Electronics Engineers (IEEE). (2024b). IEEE Std 2894-2024: IEEE Guide for an Architectural Framework for Explainable Artificial Intelligence. IEEE. \url{https://standards.ieee.org/ieee/2894/11296/}}

\refentry{ISO/IEC. (2023). ISO/IEC 42001:2023 Information technology - Artificial intelligence - Management system. International Organization for Standardization. \url{https://www.iso.org/standard/42001}}

\refentry{Mitchell, M., Wu, S., Zaldivar, A., Barnes, P., Vasserman, L., Hutchinson, B., Spitzer, E., Raji, I. D., \& Gebru, T. (2019). Model Cards for Model Reporting. Proceedings of the Conference on Fairness, Accountability, and Transparency (FAT '19), 220-229. \url{https://doi.org/10.1145/3287560.3287596}}

\refentry{Mökander, J., \& Floridi, L. (2021). Ethics-Based Auditing to Develop Trustworthy AI. Minds and Machines, 31, 323-327. \url{https://doi.org/10.1007/s11023-021-09557-8}}

\refentry{Nannini, L., Smith, A. L., Maggini, M. J., Panai, E., Feliciano, S., Tiulkanov, A., Maran, E., Gealy, J., \& Bisconti, P. (2026). AI Agents Under EU Law. arXiv preprint arXiv:2604.04604. \url{https://arxiv.org/abs/2604.04604}}

\refentry{National Information Security Standardization Technical Committee (TC260). (2025). Artificial Intelligence Safety Governance Framework (Version 2.0). Released September 2025. Cyberspace Administration of China.}

\refentry{Organisation for Economic Co-operation and Development (OECD). (2024). Recommendation of the Council on Artificial Intelligence (updated May 2024). OECD Legal Instruments, OECD/LEGAL/0449. \url{https://legalinstruments.oecd.org/en/instruments/OECD-LEGAL-0449}}

\refentry{OWASP Foundation. (2025, December 9). OWASP Top 10 for Agentic Applications 2026. OWASP GenAI Security Project. \url{https://genai.owasp.org/resource/owasp-top-10-for-agentic-applications-for-2026/}}

\refentry{Pandit, H. J., \& Golpayegani, D. (Eds.). (2026). AI Technology Concepts (Data Privacy Vocabulary v2.3 AI Extension). W3C Data Privacy Vocabularies and Controls Community Group Final Report, February 2026. \url{https://w3id.org/dpv/2.3/ai}}

\refentry{Raji, I. D., Smart, A., White, R. N., Mitchell, M., Gebru, T., Hutchinson, B., Smith-Loud, J., Theron, D., \& Barnes, P. (2020). Closing the AI Accountability Gap: Defining an end-to-end framework for internal algorithmic auditing. Proceedings of the 2020 Conference on Fairness, Accountability, and Transparency (FAT '20), 33-44. \url{https://doi.org/10.1145/3351095.3372873}}

\refentry{Slattery, P., Saeri, A. K., Grundy, E. A. C., Graham, J., Noetel, M., Uuk, R., Dao, J., Pour, S., Casper, S., \& Thompson, N. (2026). The AI risk repository: A meta-review, database, and taxonomy of risks from artificial intelligence. Patterns. \url{https://doi.org/10.1016/j.patter.2026.101517}}

\refentry{Tabassi, E. (2023). Artificial Intelligence Risk Management Framework (AI RMF 1.0) (NIST AI 100-1). National Institute of Standards and Technology. \url{https://doi.org/10.6028/NIST.AI.100-1}}

\refentry{Wang, B., Chen, W., Pei, H., Xie, C., Kang, M., Zhang, C., Xu, C., Xiong, Z., Dutta, R., Schaeffer, R., Truong, S. T., Arora, S., Mazeika, M., Hendrycks, D., Lin, Z., Cheng, Y., Koyejo, S., Song, D., \& Li, B. (2023). DecodingTrust: A Comprehensive Assessment of Trustworthiness in GPT Models. Advances in Neural Information Processing Systems 36 (NeurIPS 2023), Datasets and Benchmarks Track. arXiv:2306.11698. \url{https://arxiv.org/abs/2306.11698}}

\refentry{Zeng, Y., Klyman, K., Zhou, A., Yang, Y., Pan, M., Jia, R., Song, D., Liang, P., \& Li, B. (2024). AI Risk Categorization Decoded (AIR 2024): From Government Regulations to Corporate Policies. arXiv preprint arXiv:2406.17864. \url{https://arxiv.org/abs/2406.17864}}

\end{document}